\def\eeq{\relax}
\def\beq#1#2\eeq{\begin{equation}\label{#1}#2\end{equation}}
\def\bal#1#2\eal{\begin{align}\label{#1}#2\end{align}}
\def\bse#1#2\ese{\begin{subequations}\label{#1}#2\end{subequations}}
\def\ba{\begin{aligned}}
\def\ea{\end{aligned}}

\def\div{\operatorname{div}}

\def\Div{\operatorname{Div}}
\def\det{\operatorname{det}}
\def\dd{\operatorname{d}}
\def\tr{\operatorname{tr}}

\def\cS{{\cal Q}}
\def\g{^{(0)}}

\newcommand{\ga}{\gamma}

\newcommand{\de}{\delta}

\newcommand{\vareps}{\varepsilon}
\newcommand{\la}{\lambda}

\newcommand{\om}{\omega}

\newcommand{\si}{\sigma}

\newcommand{\pa}{\partial}

\newcommand{\been}{\begin{equation}}
\newcommand{\een}{\end{equation}}
\newcommand{\beena}{\begin{eqnarray}}
\newcommand{\eena}{\end{eqnarray}}

\newcommand{\tn}{\textnormal}

\documentclass[11pt]{article}

\usepackage{amssymb,mathtools}
\usepackage{graphicx}
\usepackage{subfigure}
\usepackage{psfrag}
\usepackage{float}

\usepackage{color}

\renewcommand{\appendix}{
  \setcounter{section}{0}\renewcommand{\thesection}{\Alph{section}}
  \section*{Appendix} }

\def\Appendix#1{
  \setcounter{equation}{0}
  \renewcommand{\theequation}{\thesection.\arabic{equation}}
  \section{#1}  }

\begin{document} 
\def\singlespacing{\baselineskip=13pt}	\def\doublespacing{\baselineskip=18pt}
\singlespacing

\title{Hyperelastic cloaking theory: Transformation elasticity \\ with pre-stressed solids}

\author{A. N. Norris$^*$ and W. J. Parnell$^\dagger$
\\ \\
$*$Mechanical and Aerospace Engineering, Rutgers University,\\
Piscataway, NJ 08854-8058, USA\\
$\dagger$School of Mathematics, Alan Turing Building, \\
University of Manchester, Manchester, M13 9PL, UK.
}

\maketitle

\begin{abstract}

Transformation elasticity, by analogy with transformation acoustics and  optics, converts material  domains  without altering  wave properties, thereby enabling cloaking and related effects.     By noting  the similarity between  transformation elasticity and  the theory of  incremental motion superimposed on finite pre-strain it is shown that
 the constitutive parameters of transformation elasticity  correspond to the  density and moduli of  small-on-large theory.  The formal equivalence indicates that   transformation  elasticity  can be achieved by  selecting a particular   finite (hyperelastic) strain energy function, which for   isotropic elasticity is  semilinear  strain energy.   The associated elastic transformation  is restricted by the requirement  of statically equilibrated pre-stress.  This constraint can be cast as $\tr {\mathbf F} =$ constant, where $\mathbf{F}$ is the deformation gradient, subject to symmetry constraints, and  its consequences are explored both analytically and  through numerical examples of cloaking of anti-plane and in-plane wave motion.

\end{abstract}


\section{Introduction}\label{sec1}

The  principle underlying cloaking of electromagnetic and acoustic waves  is the   transformation or change-of-variables method \cite{Greenleaf03,Pendry06} whereby  the material properties of the cloak are defined by a spatial transformation.    While the first applications were to electromagnetism, e.g.\ \cite{Schurig06},   it was quickly realized that the same mathematical
methods work equally well in acoustics  \cite{Cummer07,Chen07,Cummer08}.  The fundamental identity  underlying electromagnetic and acoustic transformation is the  observation that  the Laplacian in the original coordinates maps to a differential operator  in the physical coordinates that involves a tensor which can be interpreted as the new, transformed, material properties  \cite{Greenleaf07}.    The
equivalence between the  Laplacian in the original coordinates and the new operator involves
an arbitrary divergence free tensor \cite{Norris08b}, implying for the acoustic case that the transformed material properties are not unique.  For a given transformation function one can achieve acoustic cloaking by a variety of materials, ranging from  fluids with anisotropic inertia, to quasi-solids with isotropic density but anisotropic stiffness \cite{Norris08b,Norris09}.  Non-uniqueness of the material properties  does not apply in the electromagnetic case, where, for instance, the   permittivity  and   permeability tensors  must be proportionate for a transformation of the vacuum.

A crucial aspect of transformation optics and acoustics is that the mapped  properties correspond to exotic material properties far removed from the realm of the original material.   This aspect  is accentuated in  transformation elasticity.  In the first study of
transformation theory to  elastodynamics,   Milton et al.\  \cite{Milton06} concluded that the  transformed materials are described by the Willis model.
This  constitutive theory for material response  is dispersive, involving coupling between stress and velocity, in addition to anisotropic inertia \cite{Willis97,Milton07}.
     Brun et al.\ \cite{Brun09} considered the transformation of isotropic elasticity in cylindrical coordinates and  found  transformed material properties   with  isotropic inertia and  elastic behavior of Cosserat type.  The governing equations for Cosserat elastic materials \cite{Cosserat} are the same as those of ``normal" linear elasticity except that
the  elastic moduli do not satisfy  the minor symmetry, i.e.\ $C_{jikl}^\text{eff} \ne C_{ijkl}^\text{eff}$ (although they do satisfy the major symmetry $C_{klij}^\text{eff} = C_{ijkl}^\text{eff}$).  This implies a non-symmetric stress, ${\boldsymbol \sigma}^t \ne {\boldsymbol \sigma}$ which depends not only on the strain ${\mathbf e}$ (the  symmetric part of the displacement gradient) but also on the local rotation
$ \frac12( \nabla {\mathbf u} -(\nabla {\mathbf u})^t)$.

A thorough analysis of  transformation theory for elasticity  \cite{Norris11a} indicates that, as in  acoustics, the range of mapped material properties is highly non-unique, thus explaining
the divergence in the previously obtained results \cite{Milton06,Brun09}.
The  transformed  elastodynamic   constitutive parameters  may be characterized through their dependence on  (i) the transformation (mapping function) and  (ii) on the relation between the displacement fields in the two descriptions, represented by  matrices: ${\mathbf F}$, the deformation matrix,  and  ${\mathbf A}$,
respectively.   It was shown \cite{Norris11a}
  that requiring  stress to be     symmetric  implies ${\mathbf A}={\mathbf F}$ and that the material must be of Willis form,   as Milton et al.\ \cite{Milton06} found.
  Setting    ${\mathbf A}={\mathbf I}$, on the other hand,  results in  Cosserat materials
  with non-symmetric stress but isotropic density, as found by  Brun et al.\ \cite{Brun09} and by
  Vasquez et al.\ \cite{Vasquez11a}.
  An alternative approach to transformation elasticity has been proposed that employs
inextensible fibers  embedded in an elastic material \cite{Olsson11a,Olsson11}.  This has the advantage that the effective material has isotropic density and retains both the minor and major symmetries of the stiffness tensor.
    Despite the non-unique nature of transformation elasticity  the
  materials required are, in all cases, outside  the usual realms of possibility.

In this paper we consider a class of materials displaying non-symmetric stress of the type necessary to achieve elastodynamic cloaking.  Effective moduli with the major symmetry
$(C_{ijkl} = C_{klij})$ that do not display the minor symmetry  $(C_{ijkl} \neq C_{jikl})$ are found in the theory of incremental motion superimposed on finite deformation \cite{Ogden07a}.  We take advantage of the similarities between  transformation elasticity and   small-on-large motion in the presence of finite pre-strain.  The starting point is the
 formal equivalence of  the constitutive parameters of transformation elasticity  with  the  density and moduli for incremental motion after finite  prestress.  This offers the possibility for achieving  elasticity of the desired form by proper selection of the  finite (hyperelastic) strain energy function.   Such an approach has been shown to be successful  in the context of antiplane (SH) wave motion.  By using  the neo-Hookean strain energy for incompressible solids and applying a radially symmetric cylindrical pres-strain, Parnell \cite{Parnell2011} and Parnell et al. \cite{Parnell2012} showed that the resulting small-on-large equations are identically those required for cloaking of the SH wave motion.   Here we consider the more general elastic transformation problem, {including but not limited  to SH motion}.  We show that the form of the finite strain energy is restricted in form  for isotropic elasticity.  The equivalence between the transformation and the finite pre-strain also limits the type of transformation possible.  This contrasts with the acoustic and electromagnetic problems for which  the transformation is arbitrary.  The elastic transformation  is restricted in form because the  pre-stress must be statically equilibrated,  implying that the transformation must satisfy a  partial differential equation.  We show that this constraint can be cast as $\tr {\mathbf F} =$ constant, (subject to symmetry constraints), and explore its consequences both analytically and numerically.

We begin in \S\ref{sec2} with a review of transformation elasticity and of  incremental motion superimposed on finite pre-strain, emphasizing   equivalence of the theories.   The form of the finite strain energy necessary to achieve transformation elasticity is deduced in \S\ref{sec3} and in  \S\ref{sec4}
 the constraint on the deformation  for  isotropic elasticity is  derived.     \S\ref{sec5} presents a detailed example of  the type of radially symmetric finite pre-strain   possible for isotropic elasticity.  These analytical results are  extended and illustrated in \S\ref{sec6} through numerical examples of cloaking of anti-plane and in-plane wave motion.  Conclusions are presented in \S\ref{sec7}.

\section{Background equations} \label{sec2}

We first review the theory of transformation elasticity for linearly elastic solids, and then consider the separate theory for incremental deformation in finite elasticity.

\subsection{Review of transformation elasticity}
\label{sec2.1}

\subsubsection{Transformation notation}

 A transformation from  the virtual  configuration, $\Omega_0$,  to the present  configuration $\Omega$  (also known as the physical or current domain) is described by the   mapping
from  ${\boldsymbol \xi}\in \Omega_0$ to   ${\mathbf x} \in \Omega$.
Component subscripts in upper and lower case  $(I,J,\ldots , i,j,\ldots)$  are used to distinguish between explicit dependence upon ${\boldsymbol \xi}$ and    ${\mathbf x}$, and the summation convention on repeated subscripts is assumed.    The transformation or mapping is assumed to be one-to-one and invertible.  Perfect cloaking requires that the   transformation is one-to-many at a single point in $\Omega_0$.  This can be avoided by  considering near-cloaks, where, for instance, a small hole in $\Omega_0$  is mapped to a  larger hole in $\Omega$.

The transformation gradient is defined as
${\mathbf F}\g = \nabla_{\xi} {\mathbf x}$  with inverse ${{\mathbf F}\g}^{-1} = \nabla {\boldsymbol \xi}$,
or in component form
$F\g_{iI}  = \partial x_i /\partial \xi_I$,
${F\g_{iI}}^{-1}  = \partial \xi_I /\partial x_i$.
The Jacobian of the transformation is $J_0 = \det {\mathbf F}\g $.
The  infinitesimal  displacement  ${\mathbf u}^{(0)}({\boldsymbol \xi},t) $ and   stress
${\boldsymbol \sigma}^{(0)}({\boldsymbol \xi},t) $ 
are assumed to
satisfy the equations of linear elasticity  in the virtual  domain:
\beq{91}
\div_\xi {\boldsymbol \sigma}^{(0)}  =  \rho_0\ddot{\mathbf u}^{(0)} ,
\quad
 {\boldsymbol \sigma}^{(0)} = {\mathbf C}^{(0)} \nabla_\xi {\mathbf u}^{(0)}
\quad \text{in } \Omega_0 ,
\eeq
where $\rho_0$ is the  (scalar) mass density
and the elements of the elastic stiffness tensor satisfy the full symmetries
$
C^{(0)}_{ IJKL}=C^{(0)}_{ JIKL}$, $C^{(0)}_{IJKL}=C^{(0)}_{KLIJ}$;
the first identity expresses the symmetry of the stress
and the second is the consequence of an assumed strain energy density function.

Particle displacement in the transformed domain,  ${\mathbf u}({\mathbf x},t)$,
is assumed to be related to the displacement in the virtual domain
 by the non-singular  matrix ${\mathbf A}$ as
\beq{2.1}
{\mathbf u}^{(0)} = {\mathbf A}^t {\mathbf u} \qquad (u^{(0)}_I=A_{iI}u_i).
\eeq
The choice of the transpose, ${\mathbf A}^t$ in eq.\ \eqref{2.1}, means that  the ``gauge"     ${\mathbf A}$ and the transformation gradient  ${\mathbf F}\g$  are similar objects, although at this stage they are unrelated.
 Neither ${\mathbf A}$ or ${\mathbf F}\g$ are   second order tensors because of the fact that they each have   one ``leg'' in both domains.
 Milton et al.\  \cite{Milton06}  specify ${\mathbf A}={\mathbf F}\g$ since this is the only choice that guarantees a symmetric stress (see \S \ref{333}).
Identifying \cite{Milton06}
$\dd {\boldsymbol \xi }$ and $\dd {\mathbf x}$ with ${\boldsymbol  u}\g$ and ${\boldsymbol u }$, respectively, and using
$
\dd {\boldsymbol \xi } = {{\mathbf F}\g}^{-1} \dd {\mathbf x}
$ would
lead one to expect ${\mathbf A}^t={{\mathbf F}\g}^{-1}$.  However, the displacements are not associated with the coordinate transformation and  $ {\mathbf F}\g$ and $ {\mathbf A}$ are independent quantities.

\subsubsection{The transformed equations of elasticity  }\label{333}

Under the transformation (or change of coordinates)  $ {\boldsymbol \xi } \to {\mathbf x}$ the  equilibrium  and constitutive relations \eqref{91} transform to \cite{Norris11a}
\beq{4-4}
\sigma_{ij,i} = \dot{p}_j,
\quad
\sigma_{ij} =C_{ijkl}^\text{eff}u_{l,k}+S_{ijl}^\text{eff}\, \dot{u}_l,
\quad
p_l = S_{ijl}^\text{eff}\, u_{j,i} + \rho_{jl}^\text{eff}\dot{u}_j,
\quad \text{in } \Omega_0 ,
\eeq
with parameters ${\mathbf C}^\text{eff}$, ${\mathbf S}^\text{eff}$ and ${\boldsymbol \rho}^\text{eff}$ defined as follows in the Fourier time domain (dependence $e^{-i\omega t}$ understood but omitted)
\beq{-51}
\begin{aligned}
C_{ijkl}^\text{eff} &
= J_0 C^{(0)}_{ IJKL} \cS_{ijIJ} \cS_{klKL},
\\
S_{ijl}^\text{eff} &
= (-i\omega )^{-1}  J_0 C^{(0)}_{ IJKL} \cS_{ijIJ} \cS_{klKL,k},
\\
\rho_{jl}^\text{eff}&
= \rho_0  J_0^{-1} A_{jK}A_{lK}+ (-i\omega )^{-2}   J_0 C^{(0)}_{ IJKL} \cS_{ijIJ,i} \cS_{klKL,k}
,
\end{aligned}
\eeq
where   $
\cS_{ijIJ}
= J_0^{-1} F\g_{iI} A_{jJ}$.
The elastic moduli and the density  satisfy the  symmetries
\beq{03-}
C_{ijkl}^\text{eff}  = C_{klij}^\text{eff} ,
\quad
\rho_{jl}^\text{eff} = \rho_{lj}^\text{eff},
\eeq
although these are  not the full symmetries  for the Willis constitutive model
(which requires the additional ``minor" symmetry
$C_{ijkl}^\text{eff}  = C_{jikl}^\text{eff} $).
Equations \eqref{4-4}-\eqref{-51} are the fundamental result of elastic  transformation theory  \cite{Norris11a}.

The absence of the minor symmetries under the interchange of $i$ and $j$ in
$C_{ijkl}^\text{eff}$ and $S_{ijl}^\text{eff}$ of \eqref{-51} implies that the stress is generally asymmetric.  Symmetric stress is guaranteed if $\cS_{ijIJ} = \cS_{jiIJ}$, and occurs if the gauge matrix is of the form $ {\mathbf A} = \zeta {\mathbf F}\g $,
for any scalar $\zeta \ne 0$, which may be set to unity with no loss in generality.  This ${\mathbf A}$ recovers the results of Milton et al.\ \cite{Milton06} that the   transformed material is of   the Willis form.   As  noted in \cite{Milton06}, it is the only choice for ${\mathbf A}$ that yields symmetric stress.

The equations in the transformed domain, which is the physical realm,  clearly display a great deal of non-uniqueness, corresponding to a vast realm of possible material properties.  Our preference is for non-dispersive (i.e. independent of frequency)  materials, in particular, those with the least ``unusual" properties, so that they can conceivably be related to actual materials.  In this regard, isotropic density is achieved by taking the constant matrix  ${\mathbf A}$   proportional to the identity, ${\mathbf A}=\zeta {\mathbf I}$, with $\zeta =1$ without loss of generality.  In this case
${\boldsymbol \rho}^\text{eff}  = \rho^\text{eff}  {\mathbf I} $, ${\mathbf S}^\text{eff}=0$, with non-dispersive density and elastic moduli given by
\beq{81}
 \rho^\text{eff}  =  J_0^{-1} \, \rho_0 ,
\quad
 C_{ijkl}^\text{eff}
=J_0^{-1} \,  F\g_{iI} F\g_{kK} \, C^{(0)}_{ IjKl}.
\eeq
The equations of motion in the current domain are then
\beq{577}
(C_{ijkl}^\text{eff}  {u}_{l,k})_{,i} = \rho^\text{eff} \,  \ddot{u}_{j} .
\eeq
The effective moduli of \eqref{81} satisfy the major symmetry \eqref{03-}$_1$ but
$C_{ijkl}^\text{eff}  \ne C_{jikl}^\text{eff} $, indicating a non-symmetric stress.  Departure from symmetric stress is possible in  continuum theories such as  Cosserat elasticity and micropolar theories of elasticity.   Another  context admitting non-symmetric stress is the theory of small-on-large  motion, described next.

\subsection{Small-on-large theory}\label{sec2.2}

The solid material is  considered in two distinct states: first, the   reference configuration of the solid under zero strain,  $\Omega_1$, and secondly    the
current state of the material, which is again identified with $\Omega$.
The hyperelastic theory of small motion superimposed upon large  depends upon the initial
finite, i.e.\ large, static pre-strain which maps ${\mathbf X}\in \Omega_1$ to  ${\mathbf x}\in \Omega$.
The subsequent small motion is defined by the dynamic mapping
${\mathbf X} \rightarrow {\mathbf x} +\bar{\mathbf u} ( {\mathbf x}, t)$.  The following theory assumes $\bar{\mathbf u}$ and the associated strain are sufficiently small that  tangent moduli can be employed to derive the linear equations of motion for the small-on-large motion \cite{Ogden07a}.

 Towards that end we introduce the deformation gradient of the pre-strain  ${\mathbf F} = \nabla_{X} {\mathbf x}$  with inverse ${{\mathbf F}}^{-1} = \nabla {\mathbf X}$, and   Jacobian  $J = \det {\mathbf F}$.	 The polar decomposition  is ${\mathbf F} = {\mathbf R} {\mathbf U}= {\mathbf V} {\mathbf R} $, where ${\mathbf R}$
 is proper  orthogonal ($ {\mathbf R}{\mathbf R}^t = {\mathbf R}^t {\mathbf R} = {\mathbf I}$, $\det {\mathbf R}= 1$) and the
  tensors ${\mathbf U}$, ${\mathbf V}\in$ Sym$^+$ are the  positive definite solutions of
${\mathbf U}^2 = {\mathbf C} \equiv {\mathbf F}^t{\mathbf F}$, ${\mathbf V}^2 =  {\mathbf B} \equiv {\mathbf F}{\mathbf F}^t$.
The material is  assumed to be hyperelastic, implying  the existence of a strain energy function $\mathcal{W}$   per unit   volume from which  the static Cauchy pre-stress is defined as
\beq{7=}
\sigma_{ij}^\text{pre} = J^{-1} F_{i\alpha}
 { \partial \mathcal{W} }/{ \partial F_{j\alpha} }.
\eeq
The assumed dependence of $\mathcal{W}$ on the deformation ${\mathbf F}$, along with the freedom to change the current coordinate basis (which has nothing to do with transformation!) implies that $\mathcal{W}$ must depend upon ${\mathbf Q}{\mathbf F}$ for any orthogonal ${\mathbf Q}$, and taking ${\mathbf Q} = {\mathbf R}^t$ implies the  dependence
$\mathcal{W} = \mathcal{W}({\mathbf U})$.    Assuming the density in the reference configuration is $\rho_r$, the governing equations for  subsequent small-on-large motion $\bar{\mathbf u} ( {\mathbf x}, t)$  then follow from the well known theory \cite{Ogden07a} as
\beq{5=1}
({\cal A}_{0ijkl}\bar{u}_{l,k})_{,i} = \rho \bar{u}_{j,tt},
\eeq
where
\beq{5=2}
\rho  =J^{-1} \rho_r, \quad
{\cal A}_{0ijkl} = J^{-1} \, F_{i\alpha} F_{k\beta}\, {\cal A}_{\alpha j \beta l} ,
\quad
{\cal A}_{\alpha j \beta l}
= \frac{\partial^2 \mathcal{W}}{\partial F_{j\alpha}  \partial F_{l\beta} }
\quad ( = {\cal A}_{\beta l \alpha j }) .
\eeq

\section{Potential strain energy functions} \label{sec3}

Our  objective  is to find possible  hyperelastic solids, i.e.\ strain energy functions $\mathcal{W}$   such that the equations for small-on-large motion are equivalent to those required after transformation of a  material  assumed to be homogeneous with properties $\{ \rho_0 , \, C^{(0)}_{ IjKl} \}$.

The connection  between the transformation  and the small-on-large theories is made by first identifying
the displacement fields as equivalent, $\bar{\mathbf u} ( {\mathbf x},t)= {\mathbf u} ( {\mathbf x},t)$, and then requiring that the equations of motion  \eqref{577} and \eqref{5=1} are the same.   The latter is satisfied if
\beq{5=3}
\rho= \gamma \rho^\text{eff} , \quad
{\cal A}_{0ijkl}
= \gamma  C_{ijkl}^\text{eff}   ,
\eeq
for some positive constant $ \gamma$.   Hence,
\beq{012}
J^{-1} \rho_r = \gamma  J_0^{-1} \, \rho_0
 , \qquad
J^{-1} \, F_{i\alpha} F_{k\beta}\, {\cal A}_{\alpha j \beta l}
=
 \gamma  J_0^{-1} \,  F\g_{iI} F\g_{kK} \, C^{(0)}_{ IjKl}
 .
\eeq
The   reference density  $\rho_r$ can then be chosen so that $\gamma =1$, and  eq.\ \eqref{012} then   implies
that the hyperelastic material is defined by
\beq{014}
\rho_r =
\rho_0   J_0^{-1}  J
 , \qquad
{\cal A}_{\alpha j \beta l}
=
 J_0^{-1} J \,   F_{\alpha i}^{-1}    F\g_{iI} \,   F_{\beta k}^{-1}  F\g_{kK} \, C^{(0)}_{ IjKl}
.
\eeq
Equation \eqref{014}$_1$ is automatically satisfied if the transformation and the finite deformation are related in the following manner:
 \beq{0145}
{\mathbf F} =  \big( g  {\rho_r}/{\rho_0}  \big)^{1/3}\,  {\mathbf F}\g {\mathbf G}^{-1} ,
\quad g = \det  {\mathbf G},
\eeq
for some non-singular $ {\mathbf G}$.
Equation  \eqref{014}$_2$  combined with the expression for  ${\cal A}_{\alpha j \beta l} $ in eq.\ \eqref{5=2} yields a second order differential  equation for the
strain energy function,
\beq{5=4}
\frac{\partial^2 \mathcal{W}}{\partial F_{j\alpha}  \partial F_{l\beta} }
=  \bigg(  \frac{\rho_r}{g^2\rho_0}  \bigg)^{1/3}\, G_{\alpha I} G_{\beta K} C^{(0)}_{ IjK l}.
\eeq

Recall that $\rho_0$ and $C^{(0)}_{ IjK l}$ are constant, but at this stage the remaining quantities in
\eqref{5=4}, i.e. $\rho_r$ and $ {\mathbf G}$, are not so constrained.  The density in the reference configuration could be inhomogeneous, $\rho_r = \rho_r ({\mathbf X})$. In that  case \eqref{5=4} would not have a general solution for $\mathcal{W}$ unless $ {\mathbf G}$ also depends upon ${\mathbf X}$ in such a manner that the right hand side is
independent of ${\mathbf X}$.  This suggests that permitting $\rho_r$ to be inhomogeneous does not provide any
simplification, and we therefore take the reference density to be constant, although not necessarily the same as
$\rho_0$.   The quantity  ${\mathbf G}$ could, in principle, be a matrix function of ${\mathbf F}$, but this makes the integration of \eqref{5=4} difficult if not impossible.  We therefore restrict attention to constant ${\mathbf G}$.
Consideration of the important  case of isotropic elasticity in \S\ref{sec4} indicates that the degrees of freedom embodied in ${\mathbf G}$ do not provide any significant additional properties, and therefore
for the remainder of the paper we take
${\mathbf G}={\mathbf I}$, and set  $\rho_r= { \rho_0 }$ with no loss in generality.
In this  case the  solution of \eqref{5=4} such that $\mathcal{W}=0$ under zero deformation $( {\mathbf F}={\mathbf I})$
is
\beq{5=41}
\mathcal{W} = \frac1 2   (F_{j\alpha} -\delta_{j\alpha})   (F_{l\beta} -\delta_{l\beta})
\,
C^{(0)}_{ \alpha j\beta l}
.
\eeq

Equation \eqref{5=41} provides a formal solution for $\mathcal{W}$, one that is consistent with \eqref{5=4}.  However,
the dependence of $\mathcal{W}$ in \eqref{5=41} upon ${\mathbf F}$ points to  a fundamental  difficulty, since the strain energy should be a function of ${\mathbf U}$.  The two are not equal in general, unless
\beq{7=0}
{\mathbf R}  = {\mathbf I}
\quad
\Leftrightarrow
\quad
{\mathbf F}  = {\mathbf U}= {\mathbf V}.
\eeq
We henceforth assume \eqref{7=0} to be the case: that is, {\it we restrict consideration to deformations that are everywhere rotation-free}. Equation \eqref{5=41} then suggests  the following possible form  of the finite strain energy
\beq{5=43}
\mathcal{W} = \frac1 2   E_{j\alpha}     E_{l\beta}
\, 
C^{(0)}_{ \alpha j\beta l}
\quad \text{where}\ \  {\mathbf E}\equiv {\mathbf U}-{\mathbf I}.
\eeq
Although this  has realistic dependence on ${\bf U}$, it will not in general satisfy eq.\ \eqref{5=4}, i.e.\
${\partial^2 \mathcal{W}}/{\partial F_{j\alpha}  \partial F_{l\beta} } \ne C^{(0)}_{ \alpha j\beta l}$.
We return to this crucial point for isotropic elasticity in \S\ref{3.2.4} where we demonstrate that  eq.\ \eqref{5=4}
is satisfied by the isotropic form of \eqref{5=43} under  additional conditions.
	Note that the strain measure ${\mathbf E}$, which is sometimes called the {\it extension tensor},
has as conjugate stress measure ${\mathbf S}_a = \partial \mathcal{W}/\partial {\mathbf E} = \frac 12
( {\mathbf S}{\mathbf U} +{\mathbf U}{\mathbf S} )$ where ${\mathbf S} = J {\mathbf F}^{-1}
{\boldsymbol \sigma} ({\mathbf F}^t)^{-1}$ is the second Piola-Kirchhoff tensor
 \cite[\S2.5]{Dill06}.

We  restrict attention henceforth to the case of hyperelastic materials that are isotropic in the undeformed state.

\section{Isotropic elasticity}\label{sec4}

\subsection{Semilinear strain energy function}

The   initial moduli are
$C^{(0)}_{  \alpha j\beta l}  = \lambda_0 \delta_{\alpha j}\delta_{\beta l}
+\mu_0(\delta_{\alpha \beta }\delta_{jl}+\delta_{l\alpha  }\delta_{  j\beta })$
with original Lam\'e moduli  $\mu_0 >0$, $\lambda_0$ and
Poisson's ratio
$\nu = \lambda_0 /[2 (\lambda_0 +\mu_0)] \in (-1,\frac 12 )$.
We consider the isotropic version of the hyperelastic strain energy in \eqref{5=43},
\bal{5=7}
\mathcal{W}  = \frac{\lambda_0}2 (\tr  {\mathbf E} )^2
+\mu_0   \tr ( {\mathbf E})^2
= \frac {\lambda_0}2  (i_1 -3)^2
+\mu_0 \big( (i_1-1)^2 - 2(i_2 -1) \big)
  ,
\eal
with the latter expression
in terms of   two of the three invariants of ${\mathbf U}$:    $i_1 = \lambda_1 +\lambda_2  +\lambda_3 $,
$i_2 = \lambda_1 \lambda_2  +\lambda_2\lambda_3+\lambda_3\lambda_1 $
where $\lambda_1$, $\lambda_2$, $\lambda_3$ are the principal stretches of ${\mathbf U}$.
  Materials with strain energy \eqref{5=7} have been called semilinear \cite{Lurie68} because of its relative simplicity and the linear form of the Piola-Kirchhoff stress ${\mathbf T}_R$, related to the Cauchy stress by ${\boldsymbol \sigma}^\text{pre} = J^{-1} {\mathbf F}{\mathbf T}_R^t$; thus
	${\mathbf T}_R = 2 \mu_0  {\mathbf F} +
 \big( \lambda_0 (\tr  {\mathbf E} ) -2 \mu_0  \big)  {\mathbf R}  $.
		John \cite{John60} proposed the strain energy \eqref{5=7} based on  the explicit form of its complementary energy density in terms of ${\mathbf T}_R$, a property also noted by others \cite{Zubov70,Raasch75}.   The semilinear
		strain energy is a special case of the more general harmonic strain energy function \cite{John60}.
		Plane strain solutions for harmonic strain energy are reviewed in  \cite[\S5.2]{Ogden84}.   Sensenig \cite{Sensenig64} 	examined the   stability of circular tubes under internal pressure, while Jafari et al.\ \cite{Jafari84} considered both internal and external pressure loading.  The latter study has implications for the stability of the 		pre-strain developed here, see \S\ref{5.2.3}.



The  pre-stress follows from \eqref{7=} as
\beq{7-7}
{\boldsymbol \sigma}^\text{pre} = J^{-1}  \big[
\lambda_0 (i_1 - 3) {\mathbf V}+ 2\mu_0 ({\mathbf V}^2 - {\mathbf V})
\big] . 
\eeq
It is emphasized that we are restricting attention to deformations with ${\mathbf U}={\mathbf F}={\mathbf F}^t$, so that the Piola-Kirchhoff stress is also symmetric with  ${\mathbf T}_R =
 \lambda_0 (\tr  {\mathbf E} ) {\mathbf I}
+2 \mu_0    {\mathbf E}  $.
Applying the equilibrium equation for the finite deformation,
\beq{7=83}
\Div {\mathbf T}_R =0 \quad \Rightarrow
\quad
 \lambda_0 x_{\alpha ,\alpha j} + 2\mu_0 x_{j,\alpha\alpha}  =0 .
\eeq
We seek solutions with symmetric deformation gradient, $ x_{j,\alpha} =x_{\alpha , j} $,
and hence $x_{\alpha ,\alpha j} = x_{j, \alpha  \alpha }$.  Consequently
eq.\  \eqref{7=83} is  satisfied by finite deformations
satisfying either of the equivalent conditions $x_{j, \alpha  \alpha } =0$ and  $x_{\alpha ,\alpha j} = 0$.  Thus:
\beq{7=84}
 \text{if }\ x_{j,\alpha} =x_{\alpha , j} \ \ \text{then} \qquad
x_{\alpha ,\alpha j} = 0
\quad \Leftrightarrow \quad  x_{j, \alpha  \alpha } =0.
\eeq
Since the two partial differential equations in \eqref{7=84} are the same we need only seek solutions of one. Focusing on  $x_{\alpha ,\alpha j} = 0 $ we conclude that
the most general type of deformation ${\mathbf x} ({\mathbf X}) $ is described by
\beq{78}
\Div {\mathbf x}  = c \  ( = \text{constant} >0),
\qquad \text{where } \nabla_X {\mathbf x} = \big(\nabla_X {\mathbf x}\big)^T .
\eeq
The appearance of the positive constant of integration  in \eqref{78}$_1$ means that the sum of the principal stretches is fixed,
\beq{029}
\lambda_1+\lambda_2+\lambda_3 = c \ \ (i_1 = c).
\eeq
Further implications of the general solution \eqref{78} for a material that is isotropic in its undeformed state is explored in greater detail  in the next section. 
For now we note that the
 pre-stress follows from \eqref{7-7} and \eqref{029} as
\beq{7-9}
{\boldsymbol \sigma}^\text{pre} =   2\mu_0 J^{-1}  \, \Big(
{\mathbf V}^2 - {\mathbf V} + \frac{(c   -3)\nu}{1-2\nu}  {\mathbf V}\Big)
.
\eeq

\subsection{The limit of  $\nu = \frac 12$} \label{4.2}
It is of interest to consider the limit of the isotropic solution for $\nu = \frac 12$.
By assumption the prestress must remain finite.  Consequently, using eq.\ \eqref{7-9}, it becomes clear that in the limit as $\nu \to \frac 12$ the constant of integration
 $c\equiv 3$, i.e.\
\beq{791}
\begin{aligned}
&\mathcal{W} = \mu_0   \tr ({\mathbf U}^2-{\mathbf I})
,
\quad
{\boldsymbol \sigma}^\text{pre} = pJ^{-1}  \,{\mathbf V} +  2\mu_0 J^{-1}  \,
({\mathbf V}^2 -{\mathbf V}) ,
\\
&
\Div {\mathbf x} =3,
\quad  {\mathbf F}={\mathbf F}^T \,(={\mathbf U}={\mathbf V}),
\end{aligned}
\bigg\}
\quad \text{for }\
\nu = \frac 12
\eeq
where the scalar $p({\mathbf X})$ defines  the constraint reaction stress (the factor $J^{-1} $ is included for later simplification).  The latter arises from the limiting process of
$\nu \to \frac 12$ in eq.\ \eqref{7-9}, and has also been shown to be the unique form of the
 reaction stress for the constraint $\tr {\mathbf V} =3$
\cite{Beatty92}.   Note that in writing  ${\boldsymbol \sigma}^\text{pre} $ in \eqref{791} we maintain a term proportional to ${\mathbf V}$ in the second term rather than incorporating it with the constraint term.    This form is consistent with the requirement that $p = 0$ and hence ${\boldsymbol \sigma}^\text{pre} =0$ in the undeformed state ${\mathbf x} \equiv{\mathbf X}$.
The equilibrium equation for the pre-strain follows from eq.\ \eqref{791} as
$\nabla_X p + 2\mu_0   \nabla_X^2 {\bf x} = 0$, and since $ \nabla_X^2 {\bf x} =0$ (see eq.\ \eqref{7=84}) it follows that   $p=$ constant.

Several aspects  of \eqref{791} are noteworthy.
The limit of $\nu = \frac 12$
is usually associated with incompressibility, i.e.\ the constraint $J = 1$ or equivalently  $i_3\equiv \lambda_1\lambda_2\lambda_3 =1$, although the   reason underlying this identification originates in linear elasticity and is therefore by no means required.   Strictly speaking, the isochoric constraint  $i_3=1$ conserves volume under the deformation.
Here we find that $\nu = \frac 12$ implies the kinematic constraint  on the deformation that
$i_1 = \lambda_1+\lambda_2+\lambda_3 =3$.  The latter is associated with the notion of incompressibility in  linear elasticity in the form $\tr {\boldsymbol E}=0$ 
and in the present context can be viewed as a ``semilinear" feature, in keeping with the descriptor
\cite{Lurie68}
for the strain energy function \eqref{5=7}.  The kinematic condition, $\Div {\mathbf x} =3$ or equivalently
\beq{909}
 \lambda_1 +\lambda_2 +\lambda_3 = 3 \quad \big(
\tr {\mathbf V} =3\big) ,
\eeq
has been called the {\it Bell constraint} \cite{Beatty92} by virtue of the fact that Bell \cite{Bell85,Bell89} showed it to be consistent with numerous sets of data for metals in finite strain.  Solids satisfying this constraint have been called {\it Bell materials} \cite{Beatty92}.  In contrast to the  constraint
$\lambda_1\lambda_2\lambda_3 =1$  it can be shown that volume decreases for every deformation of a Bell material, and hence isochoric deformations are not possible \cite{Beatty92}.

Another feature of the $\nu = \frac 12$ limit is that the strain energy in \eqref{791} has the functional dependence
$\mathcal{W}=\mu_0 ( \lambda_1^2+\lambda_2^2+\lambda_3^2 -3)$. It is interesting to  compare this  with the strain energy for a neo-Hookean
solid, $\mathcal{W}_\text{NH}=\frac{\mu_0}{2} ( \lambda_1^2+\lambda_2^2+\lambda_3^2 - 3)$,   associated with
incompressibility (i.e.\  $i_3 =1$).  Both strain energies reduce to the incompressible form for linear elasticity, and the factor   $\frac 12$ appearing in $\mathcal{W}$ but not in  $\mathcal{W}_\text{NH}$ can be attributed to the different constraints in each case ($i_1=3$ or $i_3=1$).   Parnell \cite{Parnell2011} and Parnell et al. \cite{Parnell2012} considered neo-Hookean materials in the context of transformation elasticity for  isochoric deformation.   The present results indicates that the same form of the strain energy but with a different constraint yields a quite distinct class of volume decreasing  deformations.  This aspect will be examined further in the next section in terms of specific examples.

\subsection{Consistency condition} \label{3.2.4}
It remains to show that the semilinear strain energy  \eqref{5=7} satisfies
\beq{53}
 {\cal A}_{\alpha j \beta l}
=\lambda_0 \delta_{\alpha j}\delta_{\beta l}
+\mu_0(\delta_{\alpha \beta }\delta_{jl}+\delta_{l\alpha  }\delta_{  j\beta }),
\ \ \text{where} \ \
{\cal A}_{\alpha j \beta l}  \equiv \frac{\partial^2 \mathcal{W}}{\partial F_{j\alpha}  \partial F_{l\beta} }.
\eeq
Since the moduli  $\boldsymbol  {\cal A}$ are isotropic, it it sufficient to show the equivalence in any orthogonal system of coordinates.  We choose the principal coordinate system, in which the non-zero components of $\boldsymbol  {\cal A}$ for isotropic elasticity satisfy
\cite[eqs.\ (3.31)-(3.34)]{Ogden07a}
\bse{46}
\bal{46a}
{\cal A}_{iijj} &= \mathcal{W}_{ij},
\\
{\cal A}_{ijij}-{\cal A}_{ijji} &= \frac{\mathcal{W}_{i}+\mathcal{W}_{j}} {\lambda_i + \lambda_j}, \ \  i\ne j,
\label{46b}
\\
{\cal A}_{ijij}+{\cal A}_{ijji} &= \frac{\mathcal{W}_{i}-\mathcal{W}_{j}} {\lambda_i - \lambda_j}, \ \  i\ne j, \, \lambda_i \ne \lambda_j ,
\label{46c}
\\
{\cal A}_{ijij}+{\cal A}_{ijji} &=  \mathcal{W}_{ii}-\mathcal{W}_{ij} , \ \  i\ne j, \, \lambda_i = \lambda_j ,
\label{46d}
\eal
\ese
where $\mathcal{W}_{i} =\partial \mathcal{W}/\partial  \lambda_i $, $\mathcal{W}_{ij}= \partial^2 \mathcal{W}/\partial  \lambda_i\partial  \lambda_j$
, $i,j\in\{1,2,3\}$ with no summation on repeated indices.   Using $\mathcal{W}$ and $c$ as defined in eqs.\ \eqref{5=7}  and \eqref{029} gives
\beq{48}
\ba
\mathcal{W}_{i} &= \lambda_0\, (c-3) +2\mu_0\, (\lambda_i-1),
\\
\mathcal{W}_{ii} &= \lambda_0+2\mu_0 ,  \ \ \mathcal{W}_{ij} = \lambda_0 , \ \ i\ne j .
\ea
\eeq
These satisfy \eqref{46a}, \eqref{46c} and \eqref{46d}.   The remaining conditions \eqref{46b} become
\beq{49}
\mathcal{W}_{i} +\mathcal{W}_j=0 \ \
\Rightarrow \ \
 (\lambda_0+\mu_0) \, (c-3)  - \mu_0\, (\lambda_k-1)=0, \ \ i\ne j\ne k \ne i.
\eeq
Equation \eqref{49} constitutes   three conditions, which   taken together imply the unique but trivial
solution $\lambda_i = 1$, $i\in\{1,2,3\}$, i.e.\ zero pre-strain.    We avoid this by {\it restricting attention to two dimensional dynamic solutions only}, either in-plane (P/SV) or out-of-plane (SH) motion.

\subsubsection{In-plane (P/SV) motion}
The small-on-large displacements for in-plane motion are of the form
$u_1(x_1,x_2,t)$,  $u_2(x_1,x_2,t)$,  $u_3=0$.  The condition
\eqref{49} then only needs to be satisfied in the single instance $i,j=1,2$, implying that the out-of-plane extension is related to the sum of the in-plane extensions by
\beq{50}
\lambda_3 = 1 - \frac 1{2\nu} 
\,  (\lambda_1 +\lambda_2-2).
\eeq
Since $\lambda_3$ is strictly positive, this places an upper limit on the sum of the in-plane extensions:
$\lambda_1 +\lambda_2 < 2(1+\nu)$.

\subsubsection{Out-of-plane (SH) motion}
The out-of-plane SH motion is of the form $u_1=u_2=0$, $u_3(x_1,x_2,t)$.  The requirement now is that  ${\cal A}_{1313} $ and
${\cal A}_{2323} $ are both equal to $\mu_0$ in order to recover the out-of-plane equation of motion  and   associated tractions. Using \eqref{46b} and \eqref{46c}
\beq{2-8}
\ba
{\cal A}_{1313} -  \mu_0 &=
\big( \frac{\lambda_0+\mu_0}{\lambda_1+\lambda_3}\big) \,
\big[ c-3 - (1-2\nu) (\lambda_2 - 1) \big],
\\
{\cal A}_{2323} -  \mu_0 &=
\big( \frac{\lambda_0+\mu_0}{\lambda_2+\lambda_3}\big) \,
\big[ c-3 - (1-2\nu) (\lambda_1 - 1) \big] ,
\ea
\eeq
where $c$ is the constant from eq.\ \eqref{029}.
In this form it is clear that if $\nu \ne \frac 12$ then in-plane pre-stretches must be  the same,
$\lambda_1=\lambda_2 =1+ (c-3)/(1-2\nu)$, and therefore all the stretches are constant (since $c$ is a constant).  This rules out the possibility of SH cloaking since we require that the in-plane pre-strain be inhomogeneous.   However,
{\it if both} $\nu  = \frac 12 $ {\it and}
$c = 3$
simultaneously hold,  then  ${\cal A}_{1313} ={\cal A}_{2323}=  \mu_0$  for inhomogeneous and unequal in-plane stretches $\lambda_1 $ and $ \lambda_2$.
We are therefore led to the conclusion that SH cloaking requires a separate limit of the semilinear strain energy, one satisfying the Bell constraint \eqref{909} for which the strain energy and stress are given by \eqref{791}.  Note that we do not get the neo-Hookean strain energy in this limit.

\section{Applications to isotropic elasticity}\label{sec5}

\subsection{Radially symmetric cylindrical deformations }
Consider  deformations that are radially symmetric,
$r=r(R) $, $\theta=\Theta$,
in  cylindrical coordinates $(r,\theta, x_3)$ and  $(R,\Theta , X_3)$.
The stretch in the
$x_3$-direction is assumed fixed, $\lambda_3=$ constant.
The deformation gradient for $r=r(R) $  is irrotational  with
\beq{4-5}
 \big(  {\mathbf F}^t  =\big) \ \  {\mathbf F}  =
\lambda_r{\mathbf I}_r + \lambda_\theta {\mathbf I}_\theta + \lambda_3{\mathbf I}_3 ,
\quad \lambda_r = r', \ \ \lambda_\theta =\frac rR ,
\eeq
where ${\mathbf I}_r = {\boldsymbol e}_r\otimes {\boldsymbol e}_r$,
${\mathbf I}_\theta = {\boldsymbol e}_\theta\otimes {\boldsymbol e}_\theta$ and
${\mathbf I}_3 = {\boldsymbol e}_3\otimes {\boldsymbol e}_3$.
The condition \eqref{029}  implies that the sum of the in-plane
principal stretches is  constant, say $c_0$, and the  constraint \eqref{50} relates this to $c$ of eq.\ \eqref{909},
\beq{-9}
(1-2\nu) c_0 + 2\nu c = 2(1+\nu)
\quad
\text{where} \ \ c_0 = \lambda_r + \lambda_\theta.
\eeq
Equation \eqref{78} for ${\mathbf x}$   reduces to an ordinary differential equation for $r(R)$,
\beq{=10}
r '+   \frac rR = c_0,
\eeq
with  general  solution
\beq{660}
r=\frac {c_0}2 R + c_1 R^{-1}, \quad  c_1=\text{constant}.
\eeq
Note that the free parameter $c_0$ may be expressed in terms of either $c$ or $\lambda_3$, using eqs.\ \eqref{50} and
\eqref{-9}.
Using eqs.\ \eqref{7-9}, \eqref{=10}  and \eqref{4-5} it follows that the principal stretches and  stresses
for  the radially symmetric cylindrical configuration are
\beq{+2}
\ba
\lambda_r &= 2-\lambda_\theta + 2\nu (1- \lambda_3),
\ \
\lambda_\theta = \frac{r}{R},
\ \ \lambda_3,
\\
\sigma^\text{pre}_{rr} &=
  \frac{\mu_0}{\lambda_3 \lambda_\theta }
(  \lambda_r- \lambda_\theta ),
\quad
\sigma^\text{pre}_\theta  =   \frac{\mu_0}{\lambda_3 \lambda_r }
( \lambda_\theta -  \lambda_r   ) ,
\quad
\sigma^\text{pre}_{zz} &=
\frac{2\mu_0}{\lambda_r \lambda_\theta}
 (1+\nu) (\lambda_3 -1).
\ea
\eeq
Note that
\beq{4-9}
\frac rR  \to  1, \ \ \sigma^\text{pre}_{rr} \to 0 \ \text{as} \ r\to \infty \
\text{iff } \
 \lambda_3 =1  \  \ (\Leftrightarrow c_0=2).
\eeq


\subsection{Two types of cloaking}
\subsubsection{Conventional cloaking (CC)}
The conventional concept of  a cloaking material is that it occupies a finite region, in this case, the shell
$R\in [A,B]$ which maps to an equivalent shell in physical space with the same outer surface  and an inner surface
of radius larger than the original, i.e. $r\in [a,B]$, $a\in (A,B)$.  Applying  \eqref{660}  with the two constraints
$r(A)=a$ and $r(B)=B$ yields
\beq{-12}
r = R + (a-A)\bigg[
\frac{ \big(\frac BR \big)^2 - 1}{ \big(\frac BA \big)^2 - 1}
\bigg] \frac{R}{A}, \ \  R\in [A,B], \qquad \quad (\text{CC})
\eeq
which specifies the previously free parameter $\lambda_3$ (also $c$ and $c_0$) as
\beq{-31}
\lambda_3  = 1 - \frac 1{\nu}\, \bigg[
\frac{  \frac aA   - 1}{ \big(\frac BA \big)^2 - 1}
\bigg]   < 1.
\eeq
The constraint $\lambda_3  >0$ therefore sets a lower limit in the permissible value of the outer radius as
\beq{-32}
B > A\,
  \Big( 1 + \frac 1{\nu}
 \big(\frac aA   - 1  \big) \Big)^{1/2}  .
\eeq

The mapping \eqref{-12} must also be one-to-one within the shell with $\lambda_r = r' >0$.  This means that there should be no zero of  $r'=0$  for $R\in[A,B]$.  The convex nature of the solution \eqref{-12} implies there is only one zero, say at $R=R_0$.   Since sgn$\, r'= $sgn$(R-R_0)$, it follows that $R_0 <A$ must hold.  Noting from \eqref{660} that
$R_0^2=  2c_1/c_0$, and using \eqref{-12} to infer  $c_0$ and $c_1$,  the condition $R_0^2 < A^2$ becomes
\beq{83-}
  \quad
a < a_\text{max} \equiv
   {2 A}\big/ \Big[{ 1+\big(\frac AB \big)^2 }\Big]\, .
\eeq
The magnification factor  $\frac aA \ge 1 $, which measures the ratio of the radius of the mapped hole to the radius of the original one, is therefore bounded according to
\beq{120}
 \frac aA <2.
\eeq
In order to achieve a reasonable degree of cloaking one expects that the magnification factor is large, so that the mapped hole corresponds to an original hole of small radius and hence small scattering cross-section.
The limitation expressed by \eqref{120} therefore  places a severe restriction on the use of the hyperelastic material as a conventional cloak.
Note that the upper limit on $a$ in \eqref{83-}  is not strictly achievable because $a = a_\text{max}$ implies $\lambda_r (A)= r' (A) = 0$ and hence the  principal stresses
$\sigma^\text{pre}_\theta , \sigma^\text{pre}_{zz}$
 become infinite at $r=a$.

The hyperelastic mapped solid  has other aspects that further diminish its attractiveness as a conventional cloaking material.   Specifically, a non-zero traction must be imposed on both the outer and inner boundaries to maintain the state of prestress.
Noting that the radial stress is
\beq{-67}
\sigma^\text{pre}_{rr} (R) = -
\frac{2\mu_0 }{\lambda_3}    \frac Rr
\Big[
\frac rR -1 - \nu ( 1-\lambda_3)
\Big] ,
\eeq
with  $r$  given in \eqref{-12}, yields
\bal{669}
\sigma^\text{pre}_{rr} (A) =
\sigma^\text{pre}_{rr} (B)  -
\frac{2\mu_0 }{\lambda_3}    \big( 1 - \frac Aa \big) ,
\qquad
\sigma^\text{pre}_{rr} (B) =
\frac{2\mu_0 }{\lambda_3} \nu (1-\lambda_3) .
\eal
The necessity of the inner traction at $r=a $  is a reasonable condition, but the requirement for an equilibrating traction at  $r=B $ is physically difficult.   One way to avoid this is to let $B\to\infty$, considered below.

\begin{figure}[H]
				\begin{center}	
				\includegraphics[width=4.0in , 					]{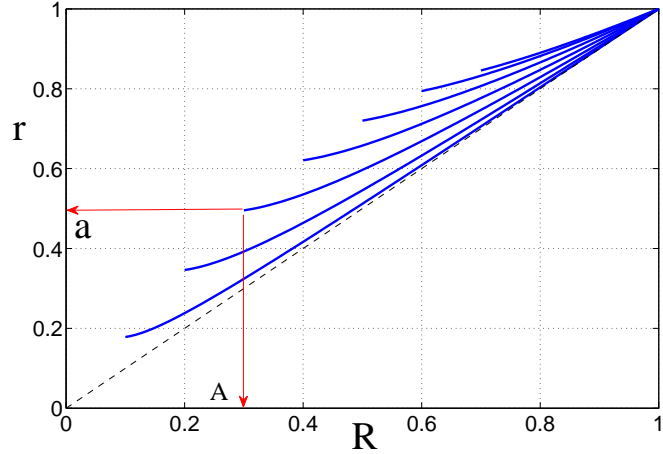}{}
	\caption{The transformed radius $r$ as a function of $R$ for 2D radially symmetric pre-strain, from eq.\ \eqref{-12}.  The seven curves correspond to
	$A=0.1, 0.2, \ldots , 0.7$ with $B=1$ in each case.  The value of the mapped inner radius
 for each curve is
	$a\equiv r(A)=0.9 \, a_\text{max}	$ where $a_\text{max}	$ is defined in eq.\ \eqref{83-}.   The dashed line indicates $r=R$.  Mappings that lie above this line represent spatial compression, applicable to cloaking. }
		\label{fig1} \end{center}
	\end{figure}

Examples of the radial deformation are given in Figure \ref{fig1}.

\subsubsection{Hyperelastic  cloaking (HC)} \label{789}
The hyperelastic material is now considered as infinite in extent.  The cloaking effect is caused by allowing a  radially symmetric hole in the unstressed configuration to be expanded under the action of an internal pressure to become a larger hole.  We therefore require that the traction at infinity is zero, and that $r/R$ tends to unity, so that \eqref{4-9} applies.    Then setting the mapped hole radius to  $r(A)=a $ $(>A)$ implies the unique mapping
\beq{-13}
r = R + (a-A) \big(  A/R \big) , \ \  R\in [A,\infty ). \qquad \quad (\text{HC})
\eeq
This deformation is simply the limiting case of \eqref{-12} for $B\to \infty$.
Note that the restriction  \eqref{120} still applies to the magnification factor  $\frac aA$, in order to ensure  $\lambda_r >0$ for $r >a$.
The traction at the inner surface is a pressure  which follows from  \eqref{669} in the limit $B\to \infty$, $\lambda_3 \to 1$, as
\beq{-14}
\sigma^\text{pre}_{rr} (A) = -p_\text{in},
\quad \text{where}
\ \
p_\text{in} =  2\mu_0  \Big(1- \frac Aa \Big)   .
\eeq
It is interesting to note that the internal pressure is independent of the Poisson's ratio $\nu$ and it is therefore the same as $p_\text{in} $ found by  \cite{Parnell2011}  considering SH incremental motion.

\subsubsection{Stability of the pre-strain} \label{5.2.3}
Jafari et al.\ \cite{Jafari84} examined the stability of  a  finite thickness tube composed of material with  harmonic strain energy, which includes semilinear strain energy as a special case.  They showed that radially symmetric two dimensional finite deformations are stable under  interior pressurization with zero exterior pressure.   This implies that the finite pre-strain HC is stable.     The stability of the CC deformation \eqref{-12} does not appear to have been considered, and remains an open question.  However, the stability of the HC, corresponding to $B\to \infty$, means there exists  a minimum $B_\text{min}$ for which CC stability is ensured for all $B> B_\text{min}$.

\subsection{The limiting case when   $\nu = \frac 12$}

In this limit  the constraint \eqref{909} applies  and the pre-stress for the radially symmetric deformation follows
from \eqref{791}  with constant ``pressure" $p$   (see \S\ref{4.2})  as
\beq{+38a}
\sigma^\text{pre}_{rr} = \frac{  2\mu_0}{\lambda_3 \lambda_\theta}
(\lambda_r  -\gamma_0 ),
\qquad
\sigma^\text{pre}_{\theta\theta}  = \frac{  2\mu_0}{\lambda_3 \lambda_r}
(\lambda_\theta  -\gamma_0 ),
\eeq
where the value of the constant $\gamma_0 = 1 + p/(2\mu_0)$ depends on the specified boundary conditions,
and $ \lambda_r =\dd r /\dd R$, $\lambda_\theta= r/R$, with $r(R)$  given  by eq.\  \eqref{660} for $c_0\equiv 2$,
  For instance, in the case of hyperelastic cloaking as defined in \S\ref{789} we find, noting the result \eqref{4-9}, that $p=0$, yielding the same interior pressure $p_\text{in}$ as  eq.\ \eqref{-14}.
\begin{figure}[H]
\centering{
\psfrag{p}{$-p$}
\psfrag{r1}{$r_1$}
\includegraphics[scale=0.8]{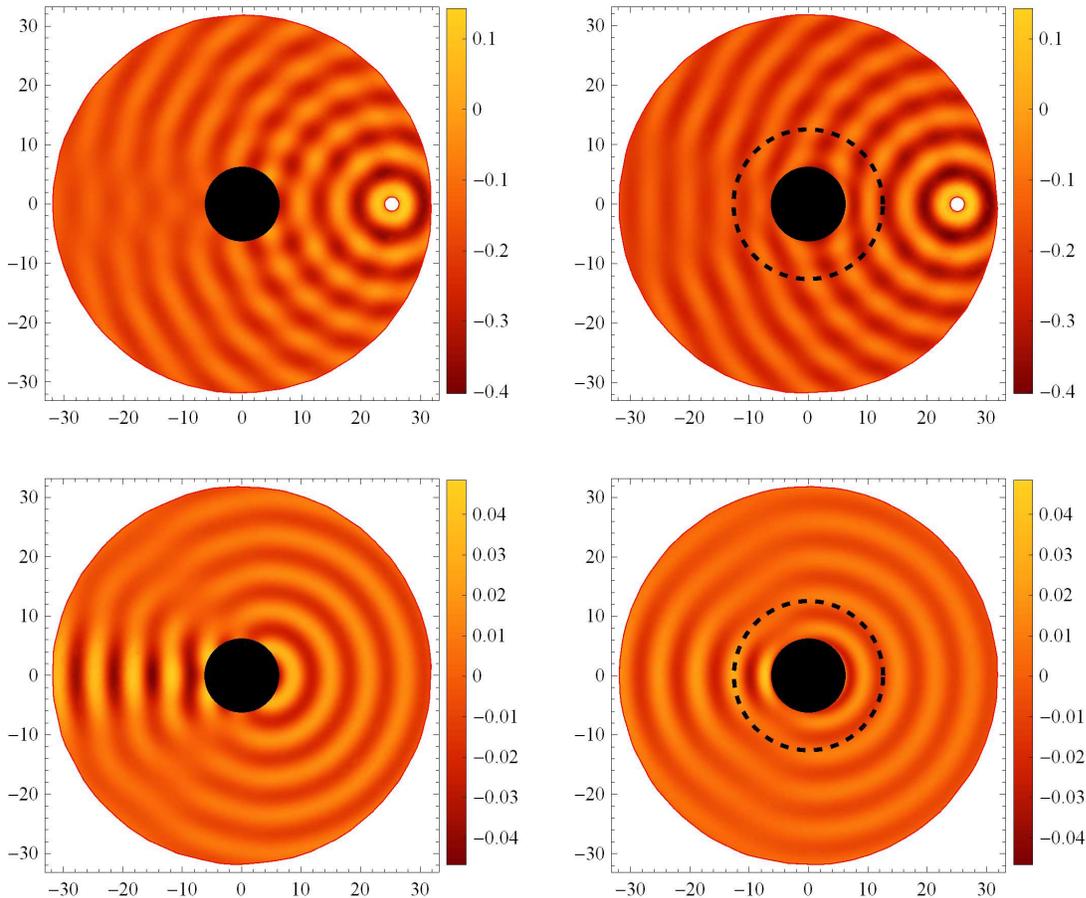}
\vspace{-.9in}
\caption{SH wave field. Left: Total (top) and scattered (bottom) fields corresponding to an undeformed cavity with scaled radius $K_sA=2\pi$. Right: Total (top) and scattered (bottom) fields corresponding to a conventional cloak generated via pre-stress where the scaled deformed inner radius is $K_sa=2\pi$ and \textit{initial} inner cavity radius defined by $a=\beta A$ where $\beta\approx 1.863$.} \label{cloak2}
}
\end{figure}

\section{Numerical examples} \label{sec6}

We illustrate the above theory in the two dimensional setting where we consider wave scattering from a cylindrical cavity with and without a cloak where the cloak is a conventional cloak created via pre-stress.  We shall show that \textit{partial cloaking} is achieved, in the sense that scattering is significantly reduced by presence of a cloak. We are not able to achieve perfect cloaking since the cavity has to be of finite radius initially and furthermore, the hyperelastic theory above restricts the expansion to be at most twice the initial radius, i.e.\ $a< 2A$. We  consider two cases: horizontally polarized shear (SH) waves and coupled compressional/in-plane shear (P/SV) waves. We take $B/a=2$ which upon using \eqref{83-} gives an initial inner to outer cloak radius ratio $B/A=1/(2-\sqrt{3})\approx=3.732$ and $a/A=\beta=1/(2(2-\sqrt{3}))\approx 1.866$. 

For the SH and P/SV wave examples considered below, we use the description in Appendices \ref{SHwaves} and \ref{PSVwaves} regarding scattering from a cylindrical cavity in an undeformed medium due to an incident field generated by a line source. In both cases considered, we assume that the line source is located at a distance $R_0$ from the centre of the cavity with $R_0/B=2$, it is of unit amplitude $C=1$ and in the P/SV case it generates purely compressional waves.

\begin{figure}
\begin{center}
\psfrag{g}{$\gamma_{SH}$}
\psfrag{p}{$\gamma_{\%}$}
\psfrag{A}{$K_sa$}
\includegraphics[scale=0.6]{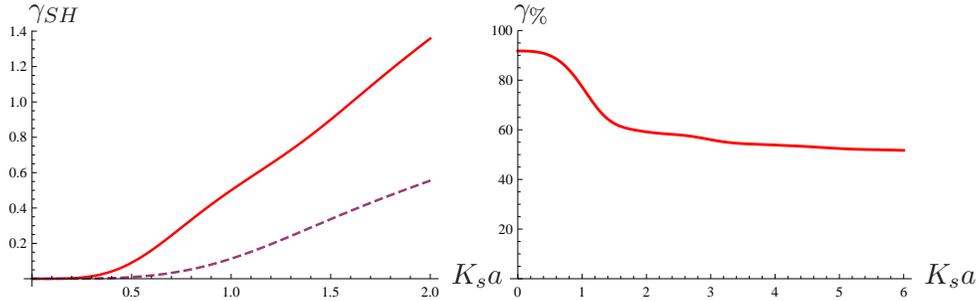}
\caption{Scaled scattering cross-section $\ga_{SH}K_s$ (left) and percentage reduction in scattering cross-section $\ga_{\%}$ by using a hyperelastic cloak (right) (with $a=\beta A$ where $\beta\approx 1.863$.), both plotted against scaled cavity radius $K_sa$ for the SH wave case. The cross-section is plotted without (solid) and with (dashed) a hyperelastic cloak. A significant reduction in scattering is achieved by using a hyperelastic cloak.} \label{SHx}
\end{center}
\end{figure}

\subsection{SH wave propagation}

In this case the shear wavenumber $K_s$ of the medium is defined by $K_s^2=\om^2/c_s^2=\rho_0\om^2/\mu_0$ where $\rho_0$ is the density of the medium in the undeformed configuration. We use the solution in Appendix \ref{SHwaves} to solve the corresponding (conventional, pre-stress) cloak problem, the difference arising merely due to the modified argument due to the hyperelastic deformation (and invariance of equations). We shall always consider the case when $R_0>B$, the outer cloak boundary. Thus in $R>B$ the solution can be written as \eqref{Wsfld},
 noting that the scattering coefficients  are equivalent to scattering coefficients for a cavity of radius $A$. Therein resides the reduction in scattering. In $a<R<B$, the total field is given by $W_i+W_s$ but with an argument given by
\begin{align}
R(r) = c_0^{-1}(r + \sqrt{r^2-2c_0c_1 }) \label{Rrdef}   
\end{align}
i.e.\ that corresponding to the hyperelastic deformation described above (see eq.\ \eqref{660}).

We take $30$ terms in the modal sum \eqref{Wsfld} for the wave field, sufficient for convergence of the solution. Figure \ref{cloak2}   shows both the total (top) and scattered (bottom) fields corresponding to the following problems: scattering from a cavity of radius $A$ with $K_sA=2\pi$ in an undeformed medium (left) and scattering from a cavity with the presence of a hyperelastic cloak (right) with undeformed ($A$) and deformed ($a$) inner radii defined via $a=\beta A$ where $\beta$ is defined above. The outer cloak boundary $B$ is defined by $K_sB=4\pi$ (right).  This demonstrates significantly reduced  scattering due to the presence of the hyperelastic cloak as compared with the non-cloaked case. Indeed we are able to quantify this by determining the reduction in scattering cross-section, defined in \eqref{SHXeqn} for plane wave incidence. Without the cloak we have $\ga_{SH} K_s= 5.39$ whereas with the cloak $\ga_{SH} K_s= 2.61$ resulting in a $51.5\%$ reduction in scattering. Figure \ref{SHx} shows the scattering cross-section $\ga_{SH} K_s$ (left) together with the percentage reduction in scattering (right).

\begin{figure}[H]
\centering{
\psfrag{p}{$-p$}
\psfrag{r1}{$r_1$}
\includegraphics[scale=0.8]{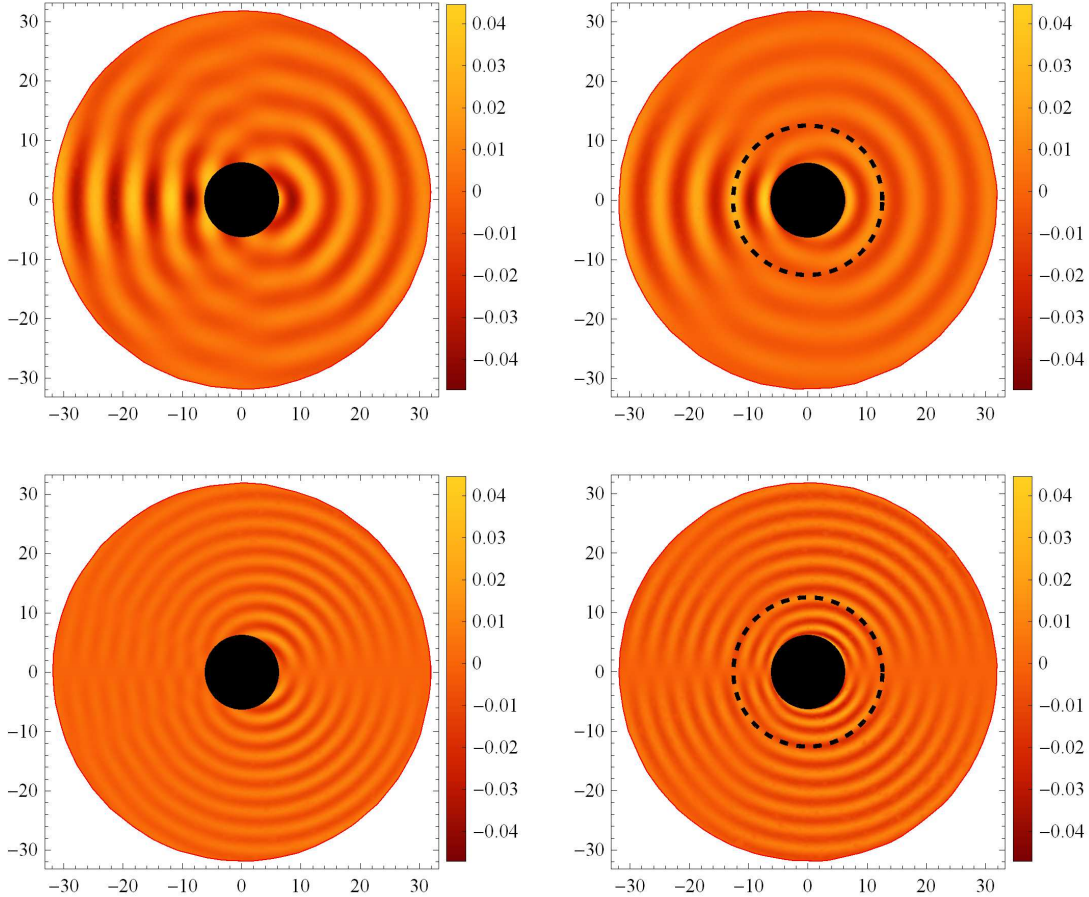}
\vspace{-.9in}
\caption{Scattered fields for the in-plane P/SV problem for an incident field generated by a compressional source at $R_0=8\pi, \Theta_0=0$. Left: Compressional (top) and shear (bottom) fields corresponding to an undeformed cavity with scaled radius $K_pA=2\pi$. Right: Compressional (top) and shear (bottom) fields corresponding to a conventional cloak generated via pre-stress where the scaled deformed inner radius is $K_pa=2\pi$ and \textit{initial} inner cavity radius is $K_pA=\pi$ so that $a=\beta A$ where $\beta\approx 1.863$.} \label{cloak3}
}
\end{figure}

\subsection{P/SV wave propagation}

In the P/SV case, in addition to the shear wavenumber $K_s$ we also introduce the compressional wavenumber $K_p$ via $K_p^2=\om^2/c_s^2=\rho_0\om^2/(\la_0+2\mu_0)$. We use the undeformed medium solution as derived in Appendix \ref{PSVwaves} as a means of determining the solution for the cloak problem. This solution is employed in the exterior region together with the same solution but with modified argument (due to the hyperelastic deformation) in the cloak region. Thus in $R>B$ the solution can be written as \eqref{PSVscat} with scattering coefficients $A_n$ and $B_n$ given by \eqref{AnPSV} and \eqref{BnPSV} respectively, noting that they are equivalent to scattering coefficients for a cavity of radius $A$ and therefore a reduction in scattering is present. Note that here a different effect is introduced as compared with the SH case: shear waves are produced as a result of mode conversion on the boundary of the cavity. In $a<R<B$, the total field is given by the sum of the scattered and incident fields but with the argument as given in \eqref{Rrdef} due to the hyperelastic deformation.

\begin{figure}[ht!]
\begin{center}
\psfrag{g}{$\gamma_{P}$}
\psfrag{p}{$\gamma_{\%}$}
\psfrag{A}{$K_pA$}
\includegraphics[scale=1.2]{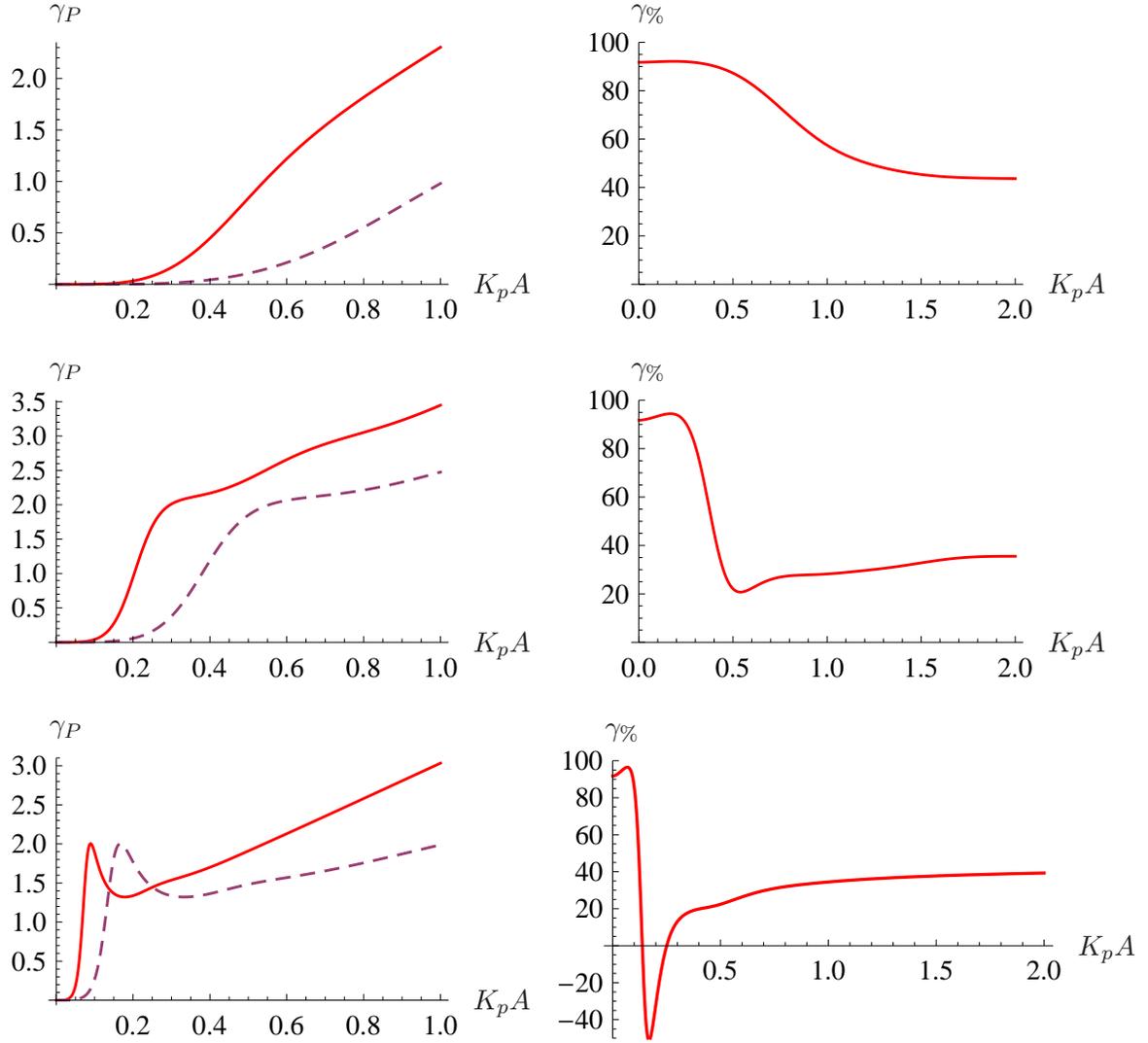}
\caption{Scattering of P/SV waves from a cylindrical cavity. Left: Scattering cross section from the undeformed cavity (solid)  with radius $K_pa$ and from a deformed cavity with initial scaled radius $K_pA$ such that $a=\beta A$ where $\beta\approx 1.863$ (dashed). Right: Percentage \textit{reduction} in scattering cross section $\gamma_{\%}$ due to pre-stress. We have $\nu=1/3$ (top), $\nu=7/15$ (middle) and $\nu=49/99$ (bottom).  Note that for the latter case the peak in scattering cross section results in a narrow range of values of $K_pa$ where the cloak \textit{increases} scattering. For other values, there is significant reduction in scattering, especially at very low frequencies.} \label{PSVx}
\end{center}
\end{figure}

We take 30 terms in the modal sums \eqref{PSVscat}, which is sufficient for convergence of the solution. Figure \ref{cloak3}   shows the scattered fields corresponding to the P-wave (top) and S-wave (bottom) fields associated with $\nu=1/3$ and for the following problems: scattering from a cavity of radius $A$ with $K_sA=2\pi$ in an undeformed medium (left) and scattering from a cavity with the presence of a hyperelastic cloak (right) with undeformed ($A$) and deformed ($a$) inner radii defined via $a/A=\beta$. The outer cloak boundary $B$ is defined by $K_sB=4\pi$ (right). Scattering is significantly reduced due to the presence of the hyperelastic cloak as compared with the non-cloaked case although it is relatively difficult to see this directly with the plots.  As with the SH case, let us quantify this by determining the reduction in scattering cross-section, defined in \eqref{PSVXeqn} for plane wave incidence. Without the cloak  $\ga_{P} K_p= 13.564$ whereas with the cloak $\ga_{P} K_p= 7.258$ resulting in a $46.48\%$ reduction in scattering. Figure \ref{PSVx} illustrates the scattering cross-section $\ga_{P} K_p$ (left) together with the percentage reduction in scattering (right)  compared for three different Poisson ratios: $\nu=1/3, 7/15$ and $49/99$. Note that for very low frequencies there is a huge reduction in scattering, close to $100\%$. This tails off at higher frequencies but still remains at around $50\%$ reduction in scattering which is clearly significant. Reduction is larger for smaller Poisson ratios. We also note the rather interesting result that the peak in the cross-section actually induces an increase in scattering at some values of $K_pa$ as compared with the case without the cloak although this is only for a narrow range of such values. This can be associated with the increasing disparity in the P and and SV wave numbers as $\nu $ tends to $\frac 12$,  noting that  $K_s^2/K_p^2 = 2(1-\nu)/(1-2\nu)$.

\section{Conclusions} 				\label{sec7}

The close correspondence between transformation elasticity and small-on-large theory points to a method for realizing the former.  Specifically, the semilinear strain energy function
of eq.\ \eqref{5=7}  yields the correct incremental moduli required for transformation of isotropic elasticity.   The connection between the two theories is that the   transformation  equals the finite deformation.   The fact that the pre-stress must be in a state of equilibrium  places a constraint on the type of transformations allowed.  Specifically, they are limited by the condition \eqref{029}, or equivalently, $\tr {\bf V} = $constant,
which yields stable radially symmetric  pre-strain \cite{Jafari84}.  This implies that the actual size of a cylindrical target can be increased in area by a factor of 4, its radius by factor of two, without any change to the scattering cross-section. The restricted form of the transformation is not surprising considering the fact that the theory can   simultaneously control more than one wave type, in contrast to acoustics.

In the two-dimensional problems for which results were provided, it was shown that the presence of a conventional cloak generated by the use of pre-stress leads to a significant reduction in the scattering cross-section from the cavity, as compared with scattering from a cavity without a cloak. This effect is particularly striking at low frequencies and for small Poisson ratios. We should note that in general one has to consider stability of nonlinear elastic solids in the large deformation regime.  While we have not undertaken  a full stability analysis, we have noted  that  the deformation for what we have termed hyperelastic cloaking (HC) is automatically stable (see \S\ref{5.2.3}).   Extension of these results will be the subject of subsequent study. We also note that manufacturing nonlinear elastic solids with specific strain energy functions can be difficult to achieve in practice, although this is certainly no harder than generating complex metamaterials which appears to be the current alternative.


This work sheds some light on  transformation methods in other wave problems.  In acoustics and electromagnetism
there is no  constraint on the transformation; any one-to-one mapping is permitted.  In principle, there is no constraint for transformation elasticity either, although the transformed materials are quite difficult if not impossible to obtain, especially since they are required to lose the minor symmetry in their corresponding elastic modulus tensor. The equivalence of transformation elasticity and small-on-large theory provides a unique and potentially realizable solution, although with a limited range of transformations allowed.  It would be desirable to relax this constraint, which interestingly, does not appear for the related problem of SH wave motion in \textit{incompressible} hyperelastic solids \cite{Parnell2011}.  The limit of incompressibility offers a clue to a possible resolution for solids with Poisson's ratio close to one-half, and will be the subject of a separate study.



\appendix

\Appendix{Elastic wave scattering from cylindrical cavities}   \label{waves}

Brief summaries of the two wave scattering problems are given below. For further details see e.g.\ Eringen and Suhubi \cite{Eri-75}. Scattering is considered from a cylindrical cavity of radius $A$, located at the origin of a Cartesian coordinate system ${\mathbf X}=(X_1,X_2,X_3)$, related to a cylindrical polar coordinate system via ${\mathbf X}=(R\cos\Theta,R\sin\Theta,Z)$. An incident wave is generated by a line source of amplitude $C$ (a force per unit length) located at the point $(R_0,\Theta_0)$. We take $\Theta_0\in[0,2\pi)$, subtended from the positive $X$ axis.

\subsection{SH wave scattering} \label{SHwaves}

In this case the line source is polarized in the $Z$ direction thus creating incident horizontally polarized shear (SH) waves which are then scattered from the cavity without mode conversion. The total wave field in this domain will therefore be $\mathbf{U}=(0,0,W(X,Y))$ where $W$ satisfies
\beq{3=2}
(\nabla^2 + K_s^2)W = { {C}{  R_0}^{-1}}\de(R-R_0)\de(\Theta-\Theta_0)
\eeq
with $K_s^2 = \rho\om^2/\mu_0$ and {$C = C_0/\mu_0$}.  We seek $W$ in the form
$W = W_i + W_s$ where $W_i= (C/4i)\tn{H}_0(K_s S)$ is the incident field and $S=\sqrt{(X-X_0)^2+(Y-Y_0)^2}$ with $X_0=R_0\cos\Theta_0, Y_0=R_0\sin\Theta_0$. We have defined $\tn{H}_0(K_sS)=\tn{H}_0^{(1)}(K_sS)=\tn{J}_0(K_sS)+i\tn{Y}_0(K_sS)$, the Hankel function of the first kind, noting that $\tn{J}_0$ and $\tn{Y}_0$ are Bessel functions of the first and second kind respectively, of order zero.
Together with the $\exp(-i\om t)$ time dependence in the problem, this ensures an outgoing field from the source. Graf's addition theorem allows us to write this field relative to the coordinate system $(R,\Theta)$ centred at the origin of the cavity \cite{Martin06} and we can use the form appropriate on $R=A$
in order to enforce the traction free boundary condition $\mu_0 \pa W/\pa R = 0 $ on $R=A$, yielding the scattered field in the form
\begin{align}
W_s = \sum_{n=0}^{\infty} \vareps_n D_n\tn{H}_n(K_sR)\cos(n(\Theta-\Theta_0))
\quad\text{with} \ \
D_n = C\frac{i}{4 } \frac{\tn{J}_n'(K_sA)}{\tn{H}_n'(K_sA)}\tn{H}_n(K_sR_0).
\label{Wsfld}
\end{align}
where $\tn{H}_n$ and $\tn{J}_n$ are respectively Hankel and Bessel functions of the first kind  of order $n$. We have also defined $\vareps_0=1$, $\vareps_n =2$, $ n\geq 1$.
If we take $R_0\rightarrow\infty$ and 
$C_0 = 2i\mu_0\sqrt{2\pi K_sR_0}e^{i(\pi/4 - K_s R_0)}$,
the incident wave of unit amplitude takes the (plane-wave) form $W_i=\exp\{iK_s(X\cos\Theta_{inc}+Y\sin\Theta_{inc})\}$
where $\Theta_{inc}=\Theta_0-\pi\in[-\pi,\pi)$ is the angle of incidence subtended from the \textit{negative} $X$ axis. The scattered wave $W_s$ takes the form \eqref{Wsfld}$_1$ with
$D_n \to D_n^{(pw)} \equiv -i^n {\tn{J}'(K_sA)}/{\tn{H}_n'(K_sA)}$.
The scattering cross-section of the cylindrical cavity for  plane wave incidence
is \cite{Lew-76}
\begin{align}
\ga_{SH} &= \frac{2}{K_s}\sum_{n=0}^{\infty}\vareps_n |D_n^{(pw)}|^2 . \label{SHXeqn}
\end{align}

\subsection{P/SV wave scattering} \label{PSVwaves}

In this case the line source at $(R_0,\Theta_0)$ with amplitude {$C_0$} is a compressional source. Thus the incident field consists purely of in-plane compressional waves. Due to mode conversion, the scattered field consists of
coupled in-plane compressional (P) and vertically polarized shear (SV) waves.  The total wave field will therefore be $\mathbf{U}=(U(X,Y),V(X,Y),0)$ and
using the Helmholtz decomposition $\mathbf{U} = \nabla\Phi + \nabla\times(\Psi\mathbf{k})$, we deduce that
\begin{align}
\nabla^2\Phi + K_p^2\Phi =
{ {C}{  R_0}^{-1}} \de(R-R_0)\de(\Theta-\Theta_0), \qquad
\nabla^2\Psi + K_s^2\Psi = 0
\end{align}
where $K_p^2 = \om^2\rho/(\lambda_0+2\mu_0)$, $K_s^2=\om^2\rho/\mu_0$ and
$C=C_0 /(\la_0+2\mu_0)$. Seek the wave field in the form $\Phi = \Phi_i + \Phi_s, \Psi = \Psi_s$ where 
$\Phi_i = (C/4i)\tn{H}_0(K_p S)$ is the incident compressional wave with notation defined in Appendix \ref{SHwaves}. We satisfy the traction free ($\si_{RR}=0$, $\si_{R\Theta}=0$) boundary condition on $R=A$, by using Graf's addition theorem, and the scattered field is
\begin{align}
\Phi_s &= \sum_{n=0}^{\infty} \vareps_n A_n H_n(K_pR)\cos(n(\Theta-\Theta_0)), &
\Psi_s &= \sum_{n=0}^{\infty} \vareps_n B_n H_n(K_sR)\sin(n(\Theta-\Theta_0)). \label{PSVscat}
\end{align}
The scattering coefficients are
\bse{33-}
\begin{align}
A_n &= \frac{i}{4} \, C \tn{H}_n(K_p R_0) \,
\big[\mathcal{I}_n^{1}(K_p A)M_n^{22}(K_s A)-\mathcal{I}_n^2(K_pA)M_n^{12}(K_sA) \big] /\Delta_n ,    \label{AnPSV}
\\
B_n &= \frac{i}{4} \, C \tn{H}_n(K_p R_0) \,
\big[\mathcal{I}_n^{2}(K_p A)M_n^{11}(K_p A)-\mathcal{I}_n^1(K_pA)M_n^{21}(K_pA) \big] /\Delta_n , \label{BnPSV}
\end{align}
\ese
where
\beq{1=4}
\ba
\mathcal{I}_n^1(x) &= \big(n^2+n-\tfrac{1}{2}(K_sA)^2 \big)J_n(x)-xJ_{n-1}(x), \\
\mathcal{I}_n^2(x) &= n(n+1)J_n(x) -nxJ_{n-1}(x), \\
M_n^{11}(x)= -M_n^{22}(x) &= \big(n^2+n-\tfrac{1}{2}(K_sA)^2 \big)H_n(x)-xH_{n-1}(x), \\
M_n^{12}(x) = -M_n^{21}(x) &= -n(n+1)H_n(x) +nxH_{n-1}(x),
\\
\Delta_n &= M_n^{11}(K_pA)M_n^{22}(K_sA) - M_n^{21}(K_pA)M_n^{12}(K_sA) .
\ea
\eeq
If we take $R_0\rightarrow\infty$ together with
$C_0 = 2i(\la+2\mu_0)\sqrt{2\pi K_pR_0}e^{i(\pi/4 - K_p R_0)}$
the incident wave of unit amplitude takes the (plane-wave) form $\Phi_i=\exp\{iK_p (X\cos\Theta_{inc}+Y\sin\Theta_{inc})\}$
where $\Theta_{inc}$ is defined above in Appendix \ref{SHwaves}.
The plane wave scattered fields take the form in \eqref{PSVscat} with
$A_n,\, B_n$ $\to A_n^{(pw)},\, B_n^{(pw)}$
where the latter are given defined in \eqref{33-} under the replacement
$\frac{i}{4}   C \tn{H}_n(K_p R_0) \to -i^n$.
The scattering cross-section $\ga_P$ of the cylindrical cavity for plane compressional wave incidence (subscript $P$ indicating this fact) 
is \cite{Lew-76}
\begin{align}
\ga_{P} &= \frac{2}{K_p}\sum_{n=0}^{\infty}\vareps_n (|A_n^{(pw)}|^2+|B_n^{(pw)}|^2). \label{PSVXeqn}
\end{align}



\subsection*{Acknowledgments}
We would like to thank  Ellis Dill  for comments and suggestions.
The work of ANN was supported by the Office of Naval Research and by the National Science Foundation.

%


\end{document}